# EXPLICIT PROOF THAT ELECTROPRODUCTION OF TRANSVERSELY POLARIZED MESONS VANISHES IN PERTURBATIVE QCD

Pervez Hoodbhoy*

*Department of Physics*

*University of Maryland*

*College Park, Maryland 20742.*

(UMD PP#02-007    August 2001)

## Abstract

By means of an explicit one-loop calculation, it is shown that the leading twist contribution to the exclusive electroproduction of transversely polarized vector mesons from the nucleon vanishes. This confirms the all-orders proof by Collins and Diehl.

Typeset using REVTEX

---

*Permanent Address: Department of Physics, Quaid-e-Azam University, Islamabad 45320, Pakistan

Electroproduction of vector mesons from a nucleon by a highly virtual longitudinally-polarized photon [2,3] is a process that is computable in perturbative QCD and, apparently, capable of accessing off-forward parton distributions [4,5]. Among others, Collins, Frankfurt and Strikman [6] showed that the reaction amplitude for the diffractive meson electroproduction can be factorized into a form involving convolutions of the off-forward parton distributions and meson wave functions with hard scattering coefficients. Of particular interest is the electroproduction of transversely polarized vector mesons because this may provide a handle to access various twist-2 chiral-odd off-forward parton distribution functions. (We refer to Refs. [7,8] for a categorization of twist-two off-diagonal distribution functions.) Experimentally, chiral-odd distributions are notoriously difficult to measure because they can make non-vanishing contributions only if matched with some other chiral-odd quantities. As noted in Ref. [6], the leading twist wave function of transversely polarized vector mesons is chirally odd. Thus there arises the possibility of accessing chiral-odd off-forward parton distributions by studying the production of transversely polarized vector mesons.

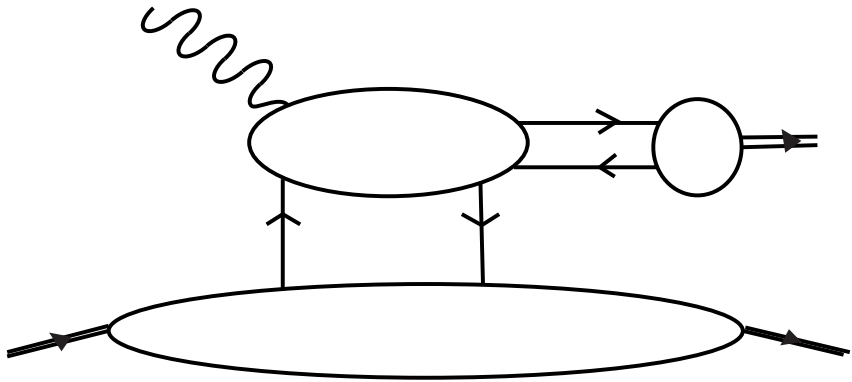

FIG. 1. Factorized vector meson electroproduction amplitude.

Unfortunately, it turns out that amplitudes associated with matching chiral-odd off-forward parton distributions with chiral-odd vector meson wave functions vanish [9] at leading order in strong coupling constant, $\alpha_s$. More interestingly, Diehl, Gousset and Pire [3] presented a proof based on chiral invariance that these hard coefficients vanish to *all orders* in perturbation theory. This proof is a direct consequence of the identity,

$$\gamma^\mu \sigma^{\rho\lambda} \gamma_\mu = 0. \tag{1}$$

Here the $\sigma^{\rho\lambda}$-matrix comes either from the density matrix associated with the off-forward quark helicity-flip distribution in the vector meson (see Fig.2a) or from that associated with the chiral-odd light-cone wave function of the nucleon (see Fig.2b), while the two $\gamma$-matrices sandwiching $\sigma^{\rho\lambda}$ correspond to the hard gluon scattering exchange. In fact, attaching any number of gluon lines to the quark lines in the basic diagram leaves the conclusion unchanged: the hard coefficients vanish to *all orders* in perturbation theory.

There may, in spite of the above, exist a way out of this conclusion. It was observed by Hoodbhoy and Lu [10] that because chiral invariance is anomalously broken in QCD, a non-zero hard coefficient may nevertheless exist. In d dimensions, the identity in Eq.(1) is replaced by,

$$\gamma^\mu \sigma^{\rho\lambda} \gamma_\mu = (d-4)\sigma^{\rho\lambda}.$$

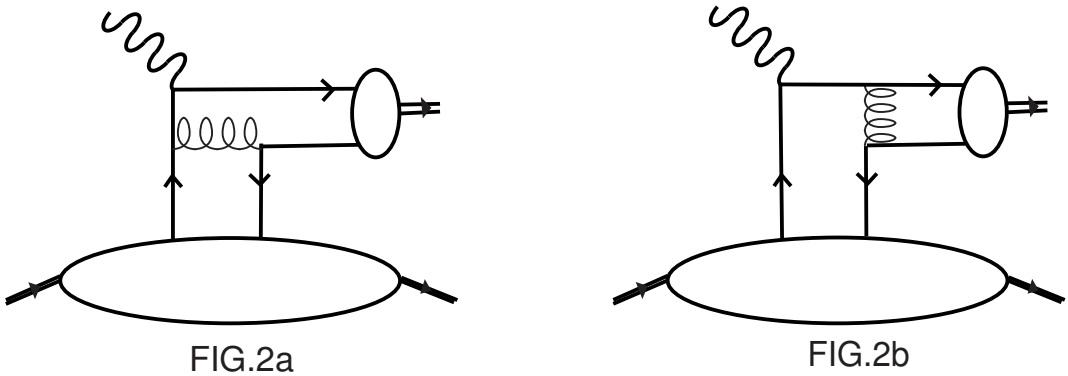

FIG. 2. Tree-level hard partonic scattering diagrams.

Thus, the Dirac trace for the quark loop will be of order $\varepsilon$, where $d = (4 + 2\varepsilon)$. Beyond the leading order, $1/\varepsilon$ divergences from poles in the loop integrals occur that cancel the factor of $\varepsilon$ in the numerator. Thus some non-vanishing terms may still survive, and this seemed to indicate that non-zero hard-scattering coefficients exist. However, it was pointed out by Collins and Diehl [1] that Hoodbhoy and Lu [10] had neglected to subtract the one-loop evolution of the nucleon and meson, which is necessary to correctly derive the hard coefficients of scattering. In this Brief Report, the above omission will be rectified and it will be seen that indeed an exact cancellation occurs between the overall amplitude and the parts coming from hadronic evolution.

Consider the process shown in Fig.1,

$$\gamma^*(q, e_L) + N(P, S) \to V(K, e_T) + N(P', S'),$$

where the first and second symbols in the parentheses stands for the particle momentum and spin vector, respectively. As usual, we define the average momentum and momentum difference for the initial and final-state nucleons:

$$\bar{P} = \frac{1}{2}(P' + P), \tag{2}$$
$$\Delta = P' - P. \tag{3}$$

It is most convenient to work in the frame in which $\bar{P}$ and $q$ are collinear with each other and put them in the third direction. In such light-cone dominated scattering processes, it is usual to introduce two conjugate light-like vectors $p^\mu$ and $n^\mu$ in the third direction with $p^2 = n^2 = 0$ and $p \cdot n = 1$. Correspondingly, using Ji's parameterization [4], the relevant momenta can be written as follows:

$$q^\mu = -2\xi p^\mu + \nu n^\mu , \tag{4}$$

$$\bar{P}^\mu = p^\mu + \frac{\bar{M}^2}{2}n^\mu , \tag{5}$$

$$\Delta^\mu = -2\xi(p^\mu - \frac{\bar{M}^2}{2}n^\mu) + \Delta_\perp^\mu , \tag{6}$$

with $\nu = Q^2/(4\xi)$, $Q^2 = -q^2$ and $\bar{M}^2 = M^2 - \Delta^2/4$. With this choice of coordinates, the longitudinal polarization vector of the virtual photon reads,

$$e_L^\mu = \frac{1}{Q}(2\xi p^\mu + \nu n^\nu) . \tag{7}$$

At lowest twist we can safely approximate the particle momenta as follows:

$$P^\mu = (1+\xi)p^\mu + \cdots , \tag{8}$$
$$P'^\mu = (1-\xi)p^\mu + \cdots , \tag{9}$$
$$K^\mu = \nu n^\mu + \cdots . \tag{10}$$

The basic idea of factorization for vector meson electroproduction is illustrated in Fig.1. According to the factorization theorem, the dominant mechanism is a single quark scattering process. The reaction amplitude is approximated as a product of three components: the hard partonic scattering, the non-perturbative matrix associated with the nucleon, and the matrix associated with the vector meson production. The active quark has to come back into the nucleon blob after experiencing the hard scattering. On the nucleon side, the initial and final quarks carry momenta $(x+\xi)p$ and $(x-\xi)p$, respectively. By decomposing the nucleon matrix one has,

$$\int \frac{d\lambda}{2\pi} e^{i\lambda x} \langle P'S'|\bar{\psi}_\alpha(-\frac{1}{2}\lambda n)\psi_\beta(\frac{1}{2}\lambda n)|PS\rangle$$
$$= \frac{\sigma_{\beta\alpha}^{\rho\lambda}}{8}[\bar{U}(P'S')H_T(x,\xi)\sigma_{\rho\lambda}U(PS) + \cdots]$$
$$\equiv \frac{i}{4}(\not{p}\not{\phi}_T)_{\beta\alpha} F_N(x,\xi,Q^2). \tag{11}$$

To save space, only one of the four twist-2 chiral-odd off-forward parton distributions have been displayed [8] in the second equation above[1]; there is no loss of generality since the remaining tensor structures have identical transformation properties. In the above, $\alpha$ and $\beta$ are the quark spinor indices. Flavor and color indices have been suppressed. Also suppressed is the gauge link operator in the definition of the matrix elements. For convenience, the calculations were performed in the Feynman gauge. On the side of the vector meson production, the collinear momenta that the quark and antiquark carry can be parameterized as $(\frac{1}{2}+z)\nu n$ and $(\frac{1}{2}-z)\nu n$ respectively. Similarly, one can write down the following decomposition for the non-perturbative matrix associated with the vector meson production,

$$\int \frac{d\lambda}{2\pi} e^{-i\lambda z} \langle 0|\bar{\psi}_\beta(-\frac{1}{2}\lambda\bar{n})\psi_\alpha(\frac{1}{2}\lambda\bar{n})|K,e_T\rangle = \frac{\sigma^{\rho\lambda}_{\alpha\beta}}{2} F_V(z,Q^2) e^*_{T\lambda} K_\rho + \cdots, \qquad (12)$$

where $F_V$ is a twist-two chiral-odd vector meson wave function.

---

[1] Recently Diehl [8] noticed that the quark and gluon helicity-flip distributions first identified in ref. [7] needed to be augmented, doubling the number of such distributions. For our purposes here, this difference is immaterial.

It is convenient to write the scattering amplitude as a perturbation series in $\alpha_s$ in the following form,

$$\mathcal{A} = \sum_{n=0} \alpha_s^n \mathcal{A}^{(n)} = \left(\frac{e}{Q}\right) \int_{-1}^{+1} dx \int_{-\frac{1}{2}}^{+\frac{1}{2}} dz \frac{F_N(x,\xi,Q^2) F_V(z,Q^2)}{(x-\xi+i\epsilon)(\frac{1}{2}-z)} \sum_{n=0} \alpha_s^n S^{(n)} \qquad (13)$$

There is no diagram at zeroth order, so $S^{(0)} = 0$.

At the tree level there are two Feynman diagrams for the hard partonic scattering as shown in Fig.2a-2b. This corresponds to the fact that either before, or after, it is struck by the virtual photon, the active quark must undergo a hard scattering to adjust its momentum so as to form the final-state vector meson. (Remember that in our chosen frame, both initial-state and final-state nucleons move in the third plus direction, while the vector meson goes in the opposite direction.) The sum of the two diagrams in Fig.2 gives the $O(\alpha_s)$ term,

$$S^{(1)} = 2C_F \pi (4-d) = 2C_F \pi (-2\varepsilon). \qquad (14)$$

Calculation of the one-loop term, of $O(\alpha_s^2)$ is more complicated. We work with renormalized perturbation theory, so all the self-energy and vertex corrections are understood to be accompanied by the corresponding ultraviolet counter-terms. It is preferable to group the diagrams by their colour structure, and the task is to calculate

$$S^{(2)} = \sum_i C_i f_i(x,\xi,z) \qquad (15)$$

where $C_i$ is the color factor and $i$ runs over all distinct one-loop Feynman diagrams, illustrated in Figs.3-5. Results for each of the different sets of diagrams is discussed next.

The diagrams shown in Fig.3 are characteristic of the three-gluon vertex and possess a common color factor of $C_{\text{fig.3}} = (N_c^2-1)/4$. After cancelling ultraviolet divergences through counter-terms, only the infrared divergences make contributions. The individual diagrams (plus their respective counterterms) yield,

$$f_{3a} = -2 - \frac{1}{(\frac{1}{2}+z)} \log(\frac{1}{2}-z) - \frac{2\xi}{\xi+x} \log\frac{\xi-x}{2\xi}, \qquad (16)$$

$$f_{3b} = -1 - \frac{1}{(\frac{1}{2}+z)} \log(\frac{1}{2}-z), \qquad (17)$$

$$f_{3c} = -1 - \frac{2\xi}{\xi+x} \log\frac{\xi-x}{2\xi}, \qquad (18)$$

$$f_{3d} = -2, \qquad (19)$$

$$f_{3e} = -2. \qquad (20)$$

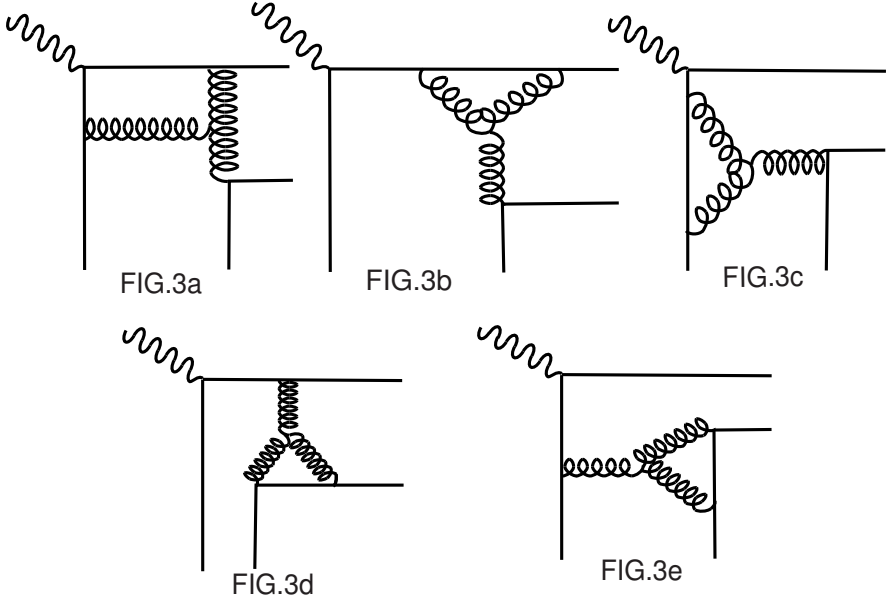

FIG. 3. One-loop corrections to the hard partonic scattering with a three-gluon vertex.

Summing over all the five diagrams in Fig.3, one has

$$\sum_{\text{fig.3}} C_i f_i(\xi, x, z) = -\frac{N_c^2 - 1}{4} \left[ 8 + \frac{2}{(\frac{1}{2} + z)} \log(\frac{1}{2} - z) + \frac{4\xi}{\xi + x} \log \frac{\xi - x}{2\xi} \right]. \qquad (21)$$

Fig.4 contains a group of diagrams that have a common color factor but vanish. The first three drop out simply because their Dirac traces vanish even in the $(4+2\varepsilon)$ space. The last two do not contribute because their vertex corrections contain no infrared divergences, while the ultraviolet divergences are canceled by the counter-terms.

Shown in Fig.5 are another group of diagrams that share the common color factor, $C_{\text{fig.5}} = -(N_c^2 - 1)/4N_c^2$. Some diagrams in this group require lengthy calculation. After considerable algebra, the contributions from individual diagrams were evaluated to be,

$$f_{5a} = -1 - \frac{\frac{1}{2} - z}{\frac{1}{2} + z} \log\left[\frac{1}{2} - z\right], \qquad (22)$$

$$f_{5b} = -1 - \frac{\xi - x}{\xi + x} \log \frac{\xi - x}{2\xi}, \qquad (23)$$

$$f_{5c} = \frac{1}{\varepsilon_I} - 2 + \log\left[\frac{(\frac{1}{2} - z)(\xi - x)}{2\xi} \frac{Q^2 e^\gamma}{4\pi\mu^2}\right],$$

$$f_{5d} = \frac{1}{\varepsilon_I} - 2 + \log\left[\frac{(\frac{1}{2} - z)(\xi - x)}{2\xi} \frac{Q^2 e^\gamma}{4\pi\mu^2}\right], \qquad (24)$$

$$f_{5e} = -\frac{1}{\varepsilon_I} - \frac{1}{\frac{1}{2} + z} \log(\frac{1}{2} - z) - \log\left[\frac{(\frac{1}{2} + z)^2(\xi - x)}{2\xi} \frac{Q^2 e^\gamma}{4\pi\mu^2}\right], \qquad (25)$$

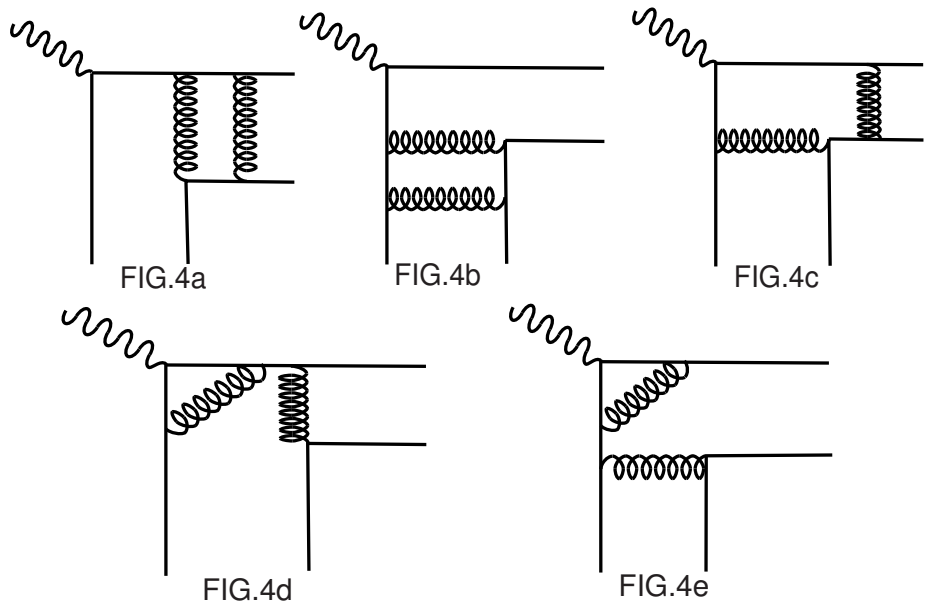

FIG. 4. A group of one-loop diagrams that vanish individually.

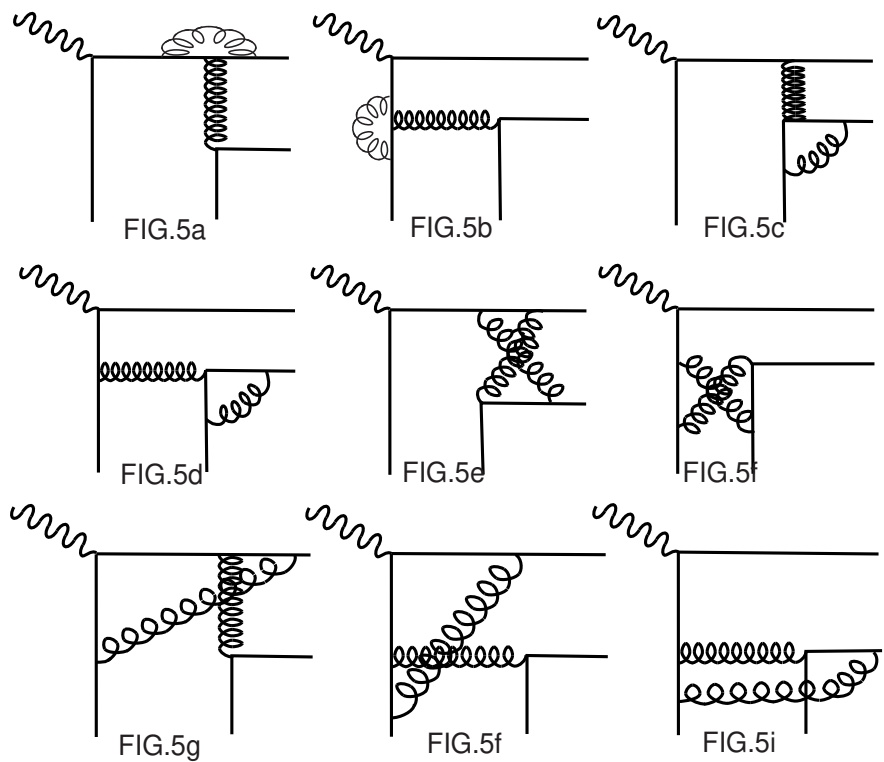

FIG. 5. A group of one-loop diagrams that share a common color factor.

$$f_{5f} = -\frac{1}{\varepsilon_I} - \frac{2\xi}{\xi+x}\log\frac{\xi-x}{2\xi} - \log\left[\frac{(\frac{1}{2}-z)(\xi+x)^2}{(2\xi)^2}\frac{Q^2 e^\gamma}{4\pi\mu^2}\right], \tag{26}$$

$$f_{5g} = +\frac{1}{\varepsilon_I} - 1 + \frac{\frac{1}{2}-z}{\frac{1}{2}+z}\log(\frac{1}{2}-z) + \log\left[\frac{(z+\frac{1}{2})(\xi+x)}{2\xi}\frac{Q^2 e^\gamma}{4\pi\mu^2}\right], \tag{27}$$

$$f_{5h} = +\frac{1}{\varepsilon_I} - 1 + \frac{\xi-x}{\xi+x}\log\frac{\xi-x}{2\xi} + \log\left[\frac{(z+\frac{1}{2})(\xi+x)}{2\xi}\frac{Q^2 e^\gamma}{4\pi\mu^2}\right], \tag{28}$$

$$f_{5i} = -\frac{2}{\varepsilon_I} - \frac{\frac{1}{2}-z}{\frac{1}{2}+z}\log(\frac{1}{2}-z) - \frac{\xi-x}{\xi+x}\log\frac{\xi-x}{2\xi} - 2\log\left[\frac{(1-z)(\xi-x)}{2\xi}\frac{Q^2 e^\gamma}{4\pi\mu^2}\right] \tag{29}$$

where $1/\varepsilon_I$ is the infrared pole, $\mu^2$ is the scale parameter in the dimensional regularization scheme, and $\gamma$ is the Euler constant. Summing over all the diagrams in Fig.5 we have,

$$\sum_{\text{fig.5}} C_i f_i(\xi, x, z) = \frac{N_c^2 - 1}{4N_c^2}\left[8 + \frac{2}{\frac{1}{2}+z}\log(\frac{1}{2}-z) + \frac{4\xi}{\xi+x}\log\frac{\xi-x}{2\xi}\right]. \tag{30}$$

At this stage, we comment on the one-loop self-energy corrections for the hard scattering partonic processes. In renormalized perturbation theory, one need not consider the self-energy insertions either on the incoming or outgoing quark lines. Instead, we need to take the diagrams in Fig.2, but in $(4+2\varepsilon)$ dimensions, and include a factor of $Z_F^{-1/2} = 1 - \frac{\alpha_s}{2\pi}C_F\frac{1}{d-4}$ for each external quark line of the hard scattering part. The ultraviolet pole in $Z_F$ can be compensated by the $\varepsilon$ factor from the tree-level trace. This is exactly where the ultraviolet divergences make their contribution. Consequently,

$$\sum_{\text{tree}} C_i f_i(\xi, x, z) = 2C_F^2. \tag{31}$$

where $C_F = (N_c^2 - 1)/(2N_c)$. On the other hand, diagrams with a self-energy insertion onto an internal line do not contribute because they have no infrared divergences.

We have by now exhausted all the one-loop diagrams for the hard scattering process. Combining Eqs. (21), (30) and (31), the following compact expression for the scattering amplitude upto $O(\alpha_s^2)$ emerges:

$$S^{(0)} + S^{(1)} + S^{(2)} = -4\alpha_s C_F \pi \varepsilon - 2\alpha_s^2 C_F^2 \left[3 + \frac{\log(\frac{1}{2}-z)}{(\frac{1}{2}+z)} + \frac{2\xi}{\xi+x}\log\frac{\xi-x}{2\xi}\right]. \tag{32}$$

Let us finally turn to the issue of factorization and extraction of the hard scattering coefficient. Schematically, the result of the above calculation for the amplitude can be written as,

$$\mathcal{A} = H * F_N * F_V. \tag{33}$$

where $H$ is the hard-scattering function and $*$ denotes convolution. We have already perturbatively evaluated $\mathcal{A}$ up to $O(\alpha_s^2)$. Each of the 3 quantities on the rhs of the above equation can be expanded as well:

$$H = H^{(0)} + \alpha_s H^{(1)} + \alpha_s^2 H^{(2)} + \cdots \tag{34}$$
$$F_N = F_N^{(0)} + \alpha_s F_N^{(1)} + \cdots \tag{35}$$
$$F_V = F_V^{(0)} + \alpha_s F_V^{(1)} + \cdots \tag{36}$$

The hard coefficients $H^{(n)}$ are to be regarded as unknowns, to be determined by substituting the power series for $\mathcal{A}$, $H$, $F_N$, and $F_V$ into Eq. (33).

From Ref. [7] (see also [11]), the first-order modification of the nucleon distribution is[2],

$$F_N^{(1)}(x,\xi,Q^2) = \frac{1}{2\pi\varepsilon}C_F \left[ \frac{3}{2} + \int_\xi^x \frac{dy}{y-x-i\epsilon} + \int_{-\xi}^x \frac{dy}{y-x-i\epsilon} \right] F_N^{(0)}(x,\xi,Q^2) +$$
$$\frac{1}{2\pi\varepsilon}C_F \left[ \theta(x-\xi)\int_x^1 dy\frac{x-\xi}{y-\xi} + \theta(x+\xi)\int_x^1 dy\frac{x+\xi}{y+\xi} \right.$$
$$\left. - \theta(\xi-x)\int_{-1}^x dy\frac{x-\xi}{y-\xi} - \theta(-\xi-x)\int_{-1}^x dy\frac{x+\xi}{y+\xi} \right] \frac{F_N^{(0)}(y,\xi,Q^2)}{y-x+i\epsilon}, \tag{37}$$

A similar calculation gives the first-order modification to the meson distribution,

$$F_V^{(1)}(z,Q^2) = \frac{1}{2\pi\varepsilon}C_F \left[ \frac{3}{2} + \int_{-\frac{1}{2}}^z \frac{dy}{y-z+i\epsilon} + \int_{\frac{1}{2}}^z \frac{dy}{y-z+i\epsilon} \right] F_V^{(0)}(z,Q^2) +$$
$$\frac{1}{2\pi\varepsilon}C_F \left[ \int_{-\frac{1}{2}}^z dy\frac{\frac{1}{2}-z}{(z-y-i\epsilon)(\frac{1}{2}-y)} + \int_z^{\frac{1}{2}} dy\frac{\frac{1}{2}+z}{(z-y+i\epsilon)(\frac{1}{2}+y)} \right] F_V^{(0)}(z,Q^2) \tag{38}$$

With all ingredients now present, we can complete the calculation. Since $\mathcal{A}^{(0)}$ vanishes, $H^{(0)}$ is also zero. $H^{(1)}$ is proportional to $\varepsilon$, but this is cancelled by the $\varepsilon^{-1}$ in $F_N^{(1)}$ and $F_V^{(1)}$ and we find that,

$$\alpha_s^2 H^{(1)} * F_N^{(1)} * F_V^{(0)} = -2\alpha_s^2 C_F^2 \left(\frac{e}{Q}\right) \int_{-1}^{+1} dx \int_{-\frac{1}{2}}^{+\frac{1}{2}} dz \frac{F_N^{(0)}(x,\xi,Q^2)F_V^{(0)}(z,Q^2)}{(x-\xi+i\epsilon)(\frac{1}{2}-z)} \left[ \frac{3}{2} + \frac{2\xi}{\xi+x}\log\frac{\xi-x}{2\xi} \right] \tag{39}$$

$$\alpha_s^2 H^{(1)} * F_N^{(0)} * F_V^{(1)} = -2\alpha_s^2 C_F^2 \left(\frac{e}{Q}\right) \int_{-1}^{+1} dx \int_{-\frac{1}{2}}^{+\frac{1}{2}} dz \frac{F_N^{(0)}(x,\xi,Q^2)F_V^{(0)}(z,Q^2)}{(x-\xi+i\epsilon)(\frac{1}{2}-z)} \left[ \frac{3}{2} + \frac{\log(\frac{1}{2}-z)}{(\frac{1}{2}+z)} \right]. \tag{40}$$

The sum of the above two terms precisely equals $\mathcal{A}^{(2)}$. Since,

$$\mathcal{A}^{(2)} = H^{(2)} * F_N^{(0)} * F_V^{(0)} + H^{(1)} * F_N^{(1)} * F_V^{(0)} + H^{(1)} * F_N^{(0)} * F_V^{(1)}, \tag{41}$$

it follows that the one-loop hard scattering coefficient vanishes, $\mathcal{H}^{(2)} = 0$.

In conclusion, the general proof by Collins and Diehl [1] appears to be correct and certainly holds at one-loop. However, it is also fairly complicated and the explicit calculation presented here shows exactly how it works out at leading order. For all this, it is still puzzling why the proof should work to all orders, given that no fundamental symmetry of QCD is being violated by the process under consideration. (Chiral symmetry is not fundamental!). No other process in QCD seems to share this property. The good news is that there will not be any leading twist chiral-odd contaminations in the measurement of chiral-even distributions. The bad news is that, at leading twist, it is impossible to access the chiral-odd parton distributions by means of vector meson electroproduction.

---

[2]There is a typographical error in Eq.9 of Ref. [7]. The correct expression involves $\int dy$ everywhere and not $\int \frac{dy}{y}$.

## ACKNOWLEDGMENTS

I would like to thank Wei Lu for his collaboration in the initial stages of the work, Xiangdong Ji for valuable suggestions and encouragement, and Andrei Belitsky for a discussion. This work is supported in part by funds provided by the U.S. National Science Foundation under grant no. INT 9820072.

# REFERENCES

[1] J.C.Collins and M.Diehl, Phys.Rev.D **61** 114015 2000.
[2] M.G.Ryskin, Z.Phys.C **37** 89 1993.
[3] M.Diehl, T.Gousset and B.Pire, Phys.Rev.D **59** 034023 1999.
[4] X.Ji, Phys.Rev.Lett. **78** 610 1997, J.Phys.G. **24** 1181 1998.
[5] A.V.Radyushkin, Phys.Lett.B **380** 417 1996; Phys.Lett.B **385** 333 1996; Phys.Rev.D **56** 5524 1997.
[6] J.C.Collins, L.Frankfurt and M.Strikman, Phys.Rev.D **56** 2982 1997.
[7] P. Hoodbhoy and X.Ji, Phys.Rev.D **58** 054006 1998.
[8] M.Diehl, Eur.Phys.J. C**19** 485 2001.
[9] L.Mankiewicz, G.Piller, E.Stein, M.Vanttinen and T.Weigl, Phys.Lett.B **425** 186 1998.
[10] P.Hoodbhoy and W.Lu, e-Print Archive: hep-ph/9902286.
[11] A.V.Belitsky and D.Muller, Phys.Lett.B **417** 129-140 1998.